\documentclass[12pt]{article}

\usepackage{bbold}
\usepackage{amssymb}
\usepackage{slashed}

\declareslashed{}{/}{.1}{0}{E}

\def\TT{{\rm TT}}

\def\psib{\overline\psi\hspace{-2.6mm}\phantom{\psi}}

\def\D{{\cal D}}

\def\J{{\cal J}}
\def\r{\rho}
\def\s{\sigma}

\def\sl#1{\slashed{#1}}

\def\ol#1{\overline{#1}}

\def\d{\partial}
\def\m{\mu}
\def\n{\nu}

\def\e{\epsilon}

\def\be{\begin{equation}}
\def\ee{\end{equation}}
\def\beq{\begin{equation}}
\def\eeq{\end{equation}}
\def\bea{\begin{eqnarray}}
\def\eea{\end{eqnarray}} 
\def\beqa{\begin{equation}\begin{array}{l}}
\def\eeqa{\end{array}\end{equation}}

\def\eqn#1{(\ref{#1})}

\def\eqref#1{eq.~(\ref{eq:#1})}

\def\a{\alpha}
 \def\G{{\it\Gamma}} \def\g{\gamma}

\def\L{{\it\Lambda}}

\def\w{\omega}

\def\nn{\nonumber}

\begin{document}

\thispagestyle{empty}
\begin{flushright}
\framebox{\small BRX-TH~478
}\\
\end{flushright}

\vspace{.8cm}
\setcounter{footnote}{0}
\begin{center}
{\Large{\bf 
(Dis)continuities of Massless Limits in Spin~3/2--mediated
Interactions and
Cosmological Supergravity }
    }\\[10mm]

{\sc S. Deser
and A. Waldron
\\[6mm]}

{\em\small  
Physics Department, Brandeis University, Waltham,
MA 02454, 
USA\\ {\tt deser,wally@brandeis.edu}}\\[5mm]


\bigskip

\bigskip

{\sc Abstract}\\
\end{center}

{\small
\begin{quote}

We extend to its spin~3/2 supersymmetric partner the very recent
demonstration that the massless limit of massive spin~2
exchange amplitudes can be made continuous in background AdS spaces,
in contrast to the known flat space 
discontinuities for both systems.
In an AdS background, unlike spin~2 
where the limit $m\rightarrow0$ is the massless one, 
spin~3/2 ``masslessness'' requires $m\rightarrow\sqrt{-\L/3}\,$, the
supergravity value
tuning the mass and cosmological constant that uniquely provides gauge
invariance and two helicities.
We find that continuity of the spin~3/2--mediated
exchange amplitude can be regained in two 
``massless'' limits $m\rightarrow0$ and $m\rightarrow\sqrt{-\L/3}\,$;
only the latter corresponds to cosmological supergravity.

\bigskip

\bigskip

\end{quote}
}

\newpage





\section{Introduction}

There is an old, but still surprising, discontinuity in the interaction
generated by spin~2 exchange between external, conserved sources
$T_{\m\n}(x)$ in flat space: the massless limit yields a finitely
different amplitude from the strictly massless case. This effect,
first observed in~\cite{vanDam:1970vg}, 
can be summarized by the following
simple formula:
\bea
{\cal A}=G_\a\,
\int d^4x\,\Big(T^{\m\n}\,D\,T'_{\m\n}
-\a\,T^\m{}_\m \,D\,T'^\n{}_\n\Big)\, ,\nn\\
m\equiv 0\, \Rightarrow\a=\frac{1}{2}\,  ; \qquad\
m\rightarrow 0 \Rightarrow \a=\frac{1}{3}\, ,
\label{tableau}
\eea
where $D$ is the scalar propagator and $G_\a$ the
corresponding gravitational constants. If the latter are chosen to fit
the observed Newtonian strength, {\it i.e.}, in the non-relativistic cases
where only the energy densities $T_{00}$ fail to vanish, then the
corresponding light bendings ($T'^\n{}_\n=0$) are discretely different
by 25\%, thereby observationally excluding any non-zero $m$, however 
minuscule\footnote{This conclusion holds at the linearized
level. The generalization of massive spin~2 to a non-linear 
Einstein-like form is, however, problematical~\cite{Boulware:1972}.}. 

It was then shown
in~\cite{Deser:1977ur}
that, as one would expect from supersymmetry, there is a precisely
parallel effect for a spin~3/2 field coupled to a conserved fermionic
source $j^\m$. Specifically, the analogue of~\eqn{tableau} is 
\bea
{\cal A}=g_\a\int d^4x\,\Big(
\ol j^\m S j_\m-\a\,\ol j^\m S \,\g_\m\g_\n\,j^\n
\Big)\, ,\nn\\
m\equiv 0\, \Rightarrow\a=\frac{1}{2}\,  ; \qquad\
m\rightarrow 0 \Rightarrow \a=\frac{1}{3}\, ,
\eea 
where $S$ is the spin~1/2 propagator, so the ``dictionary'' reads 
$T_{\m\n}\rightarrow j_\m$, $T^\m{}_\m\rightarrow \g.j$ and
$D\rightarrow S$.

Very recently, the entire question has been 
revived~\cite{Kogan:2000uy,Porrati:2000cp} 
for spin~2
with the unexpected result that continuity can in fact be recovered
by taking the background to be AdS rather
Minkowski. More precisely, it was found that the 
amplitude~\eqn{tableau} becomes
\be
{\cal A}= G \int d^4x\,\Big(T^{\m\n}\,D\,T'_{\m\n}
-\frac{a+1}{2a+3}\,T^\m{}_\m \,D\,T'^\n{}_\n\Big)\,
\ee
where $\L=-a\,m^2\leq0$. Here $\L$ is the (negative) cosmological
constant. Three ``massless'' limits are to be distinguished:
(1) The previous Minkowski one at $a=0$, where one sets $\L=0$ first
($\,\Rightarrow \a=1/3\,$), with the
consequent $m\rightarrow0$ jump compared with the strictly massless
case. (2) A ``cosmological limit'' $a\rightarrow\infty$, {\it i.e.}
$m$ vanishing faster than $\L$, which yields the welcome continuous
result~\cite{Kogan:2000uy,Porrati:2000cp} 
$\a=\frac{1}{2}$\, of the massless theory.
(3) An entire spectrum of other paths to the $m^2=0=\L$ origin
yielding any desired value of $\a$ in the range $1/3\leq\a\leq1/2$.

Although we seem to have replaced discontinuity by ambiguity, the
correct physical choice is $m$ vanishing first, for which the model
limits to the linearization of Einstein gravity in the
presence of $\L$, a perfectly gauge invariant system with just two
excitations (and not\footnote{Recently a study of massive spin~2 in
general gravitational backgrounds has been carried out with the result
that the correct excitation count holds for backgrounds where the
traceless Ricci tensor vanishes~\cite{Buchbinder:2000ar}.} 
$5$ or $6$, as in the problematic non-linear extensions~\cite{Boulware:1972}).
The final limit $\L\rightarrow0$ is therefore problem-free.

The results of~\cite{Kogan:2000uy,Porrati:2000cp}, 
themselves motivated by recent ``brane-world''
developments~\cite{Randall:1999ee}, 
beg for a parallel extension of the original
flat space system~\cite{Deser:1977ur} 
to AdS. We will indeed find the same
physical conclusion, as befits the fact that ${\cal N}=1$, 
$D=4$ cosmological supergravity~\cite{Townsend:1977qa}
describes a gauge invariant $s=3/2$ excitation with two degrees of
freedom, just like its graviton partner~\cite{Deser:1977uq}. There is, however,
an amusing new wrinkle here, related to the fact that the ``massless''
gravitino, unlike the graviton, in fact does have a ``mass term''. 
(Neither propagate on the null cone in AdS spaces, 
however~\cite{Deser:1983tm}.)
Consequently, the correct limit
that yields the continuous result 
is not setting the  parameter $m\rightarrow 0$ first but
rather letting 
\be
m\rightarrow\sqrt{-\L/3}\, ,
\label{hedge}
\ee
the ``tuned'' gauge invariant ({\it i.e.} locally
supersymmetric) value which we will review later in the text.
Our result is that the exchange amplitude connects smoothly to the strictly
massless flat case when one first takes the massless cosmological supergravity 
limit~\eqn{hedge} and thereafter $\L\rightarrow0$.
Interestingly enough, we also find that the limit $m\rightarrow0$,
followed by $\L\rightarrow0$ can be made to connect smoothly to the massless
amplitude\footnote{As this work was being completed,
reference~\cite{Grassi:2000dm} 
appeared. There the $m\rightarrow0$ non-gauge invariant ``de Sitter'' 
limit is presented and agrees with our result.}. It is
tempting to view this bifurcation of possible limits for spin~3/2 as
AdS and dS phenomena, respectively (recall that the
spin~2 result of~\cite{Kogan:2000uy,Porrati:2000cp} 
is valid for both spaces\footnote{For the spin~2 dS case,
see~\cite{Higuchi:1987py}.}) although the absence of a
dS supergravity theory~\cite{Pilch:1985aw} makes a precise relation
problematic. However, as we shall see, the current conservation law necessary
to derive the ``de Sitter'' result is not the usual supergravity 
one for spin~3/2 fields in AdS space. Further, for masses
$m>\sqrt{-\L/3}$, it is not possible to connect to the the
$m\rightarrow0$
regime without passing through the massless point $m=\sqrt{-\L/3}$.
We conclude, therefore, that 
the physical AdS massless limit $\;m\rightarrow\sqrt{-\L/3}$ 
uniquely resolves the spin~3/2 supergravity analogue of
the van Dam--Veltman--Zhakarov discontinuity.

\section{Spin~3/2}

The massive spin~3/2 Lagrangian coupled to gravity 
is\footnote{Our metric is ``mostly plus'' and Dirac
matrices are in turn ``mostly hermitean''. 
We (anti)symmetrize with unit weight; for details see~\cite{Deser:2000dz}.}
\begin{equation}
{\cal L}
=-\sqrt{-g}\; \psib_\m\,\g^{\m\n\r}\,\D_\n\psi_\r\, ,
\end{equation}
where the mass term has been introduced by defining
\be
\D_\m\equiv D_\m+\frac{m}{2}\,\g_\m\, ,
\ee
satisfying $[\D_\m,\D_\n]=[D_\m,D_\n]+(m^2/2)\,\g_{\m\n}$.
The covariant derivative on the Rarita--Schwinger field reads
\be
D_\m\psi_\n=
 \d_\m\psi_\n-\G^\r{}_{\m\n}\,\psi_\r
+\frac{1}{4}\,\w_{\m mn}\g^{mn}\psi_\n\, , 
\ee
with commutator
\be
[D_\m,D_\n]\,\psi_\r=
-R_{\m\n\r}{}^\s(g)\,\psi_\s+\frac{1}{4}\,R_{\m\n mn}(\w)\g^{mn}\,\psi_\r
\, .
\ee
We generally drop the labels $g$ and $\w$ with the curvature
convention 
$R_{\m\n\r\s}\equiv R_{\m\n\r\s}(g)=-e_\r{}^ae_\s{}^bR_{\m\n ab}(\w)$. 
The Rarita--Schwinger field equation is
\be
\g^{\m\n\r}\D_\n\psi_\r=0\, .
\label{RS}
\ee

In an AdS background, we have 
\be
R_{\m\n\r\s}=-\frac{2\L}{3}\,g_{\m[\r}g_{\s]\n}\, ,\qquad
[D_\m,D_\n]\,\psi_\r=\frac{2\L}{3}\,g_{\r[\m}\psi_{\n]}
+\frac{\L}{6}\,\g_{\m\n}\psi_\r\, .
\ee
Computations are vastly simplified by introducing half-integer spin
Lichnerowicz operators~\cite{Lichnerowicz:1961}
\bea
\D_L^{3/2}\,\psi_\m&\equiv&
\g_{\m\n\r}D^\n \psi^\r+\sl D\psi^\m=2\sl D\psi_\m+
\g_\m(\sl D\g.\psi-D.\psi)-D_\m\g.\psi \, \;\;\;\\
\D_L^{1/2}\,\chi&\equiv&\sl D\chi\, ,
\eea
with the useful properties
\bea
\D^{3/2}_L \,\g_\m\,\chi&=&\g_\m\, \D^{1/2}_L\,\chi\, \\
\D^{3/2}_L \,D_\m\,\chi&=&D_\m\, \D^{1/2}_L\,\chi\,  \\
\Big(\D^{1/2}_L\Big)^2  \,\chi&=&(D^2-\L)\,\chi \, .
\eea
The final ingredient we need is
the covariant trans\-verse--gamma-traceless decomposition in AdS:
\bea
\psi_\m&=&\psi_\m^{\TT}
+\g_\m\,\frac{1}{3D^2+\L}\,(D^2 \g.\psi-\sl D D.\psi)
-D_\m\,\frac{1}{3D^2+\L}\,(\sl D\g.\psi-4D.\psi)\nn\\
&\equiv&\psi_\m^{\TT}+\g_\m\phi+D_\m\chi\, , \qquad
D.\psi^{\TT}=0=\g.\psi^{\TT}\, .
\label{TT}
\eea

We will present two computations. The first treats the limit where
the parameter $m\rightarrow 0$ in a cosmological space. As stated
earlier, it does \underline{not}
correspond to the massless AdS supergravity limit but is, in some sense, the 
direct analogue of the spin~2 computation 
of~\cite{Kogan:2000uy,Porrati:2000cp}.
Recalling that the latter 
was valid in
either dS or AdS space, for lack of a better name,
we refer to this limit as the ``de Sitter'' calculation. 
The second computation takes the limit 
$m\rightarrow\sqrt{-\L/3}\,$, the massless locally
supersymmetric point, and will verify the continuity of
cosmological AdS supergravity.

We begin with the ``de Sitter'' case. 
Our method follows the lines of~\cite{Porrati:2000cp}. We compute
a single particle exchange amplitude between two covariantly 
conserved spinor-vector sources $j _\m$ and $\ol j _\m$, $D.j=0
=\ol j . \overleftarrow{D}$
\be
{\cal A}=\int d^4x\, \sqrt{-g}\;\ol j ^\m \Delta^{3/2}_{\m\n}(\L) j ^\n 
=\int d^4x\, \sqrt{-g}\;\ol j ^\m \psi_\m(j )\, ,
\ee
where $\Delta^{3/2}_{\m\n}(\L)$ is the propagator for a massive spin~3/2
particle in an AdS space and the second equality is achieved by substituting 
the solution $\psi_\m(j )$ of the Rarita--Schwinger equation with a source
\be
\g^{\m\n\r}\D_\n\psi_\r=j ^\m\, .
\label{RSj}
\ee
Since the source $\ol j _\m$ is covariantly conserved we only need to calculate 
$\psi_\m^\TT(j )$ and $\phi(j )$. The second of these is computed as follows:
Note that
\bea
\D.j &=&\frac{1}{2}\,m\,\g.j =
\D_\m\g^{\m\n\r}\D_\n\psi_\r=-\frac{1}{2}\,(\L+3m^2)\,\g.\psi
\label{Bianchi}\\
\g.j &=&(2\sl D-3m)\,\g.\psi-2D.\psi\, .
\eea
Solving 
and 
substituting these into~\eqn{TT} gives (note that $[\sl D,D^2]\g.j=0$)
\be
\phi(j )=\,\frac{\L(\sl D-2m)}{2\,(3D^2+\L)\,(\L+3m^2)}\;\g.j \, .
\label{phi}
\ee

Proceeding to the computation of $\psi_\m^\TT(j )$ we rewrite~\eqn{RSj}
in terms of the Lichnerowicz operator
\be
(\D^{3/2}_L-\sl D)\,\psi_\m-m\,\g_{\m\n}\psi^\n=j _\m\, .
\ee
Using $(\g_\m\phi)^{\TT}=0= (D_\m \chi)^\TT$ along with the
properties
of the Lichnerowicz operators produces the identities
\be
\Big(\D^{3/2}_L\psi_\m\Big)^\TT\!=\D^{3/2}_L\psi_\m^\TT\!
=\Big(2\sl D\,\psi_\m\Big)^\TT\!
=2\sl D\,\psi_\m^\TT
\, ,\!\quad
\Big(m\g_{\m\n}\psi^\n\Big)^\TT\!=m\,\psi^\TT_\m\, .
\label{lick}
\ee
Hence
\be
\psi_\m^\TT(j )=\frac{1}{\frac{1}{2}\,\D^{3/2}_L+m}\;j _\m^\TT
=\frac{1}{\sl D+m}\;j _\m^\TT\, .
\label{TTe}
\ee
For a covariantly conserved quantity, the transverse--gamma-traceless
decomposition simplifies;
\be
j _\m^\TT=j _\m
-(\g_\m\,D^2 - D_\m \sl D)\,  \frac{1}{3D^2+\L}\, \g.j \, .
\ee
Using this result for $\ol j _\m$ and equations~\eqn{TTe}
and~\eqn{phi} we obtain our result for the exchange
amplitude\footnote{The AdS identity
$$
\frac{1}{\sl D+m}\,(\g_\m D^2-D_\m \sl D)\,\chi=
-(\g_\m D^2-D_\m \sl D)\,\frac{\sl D+m}{D^2-m^2}\,\chi
-\frac{\L}{2}\,(\g_\m\sl D-4D_\m)\,\frac{1}{D^2-m^2}\,\chi
$$
for any spinor $\chi$
has also been used to derive this result. It follows from the
identities $\sl D^2\g_\m\chi=\g_\m D^2\chi$ and
$\sl D^2D_\m\chi=D_\m D^2\chi$.}
\bea
{\cal A}
&=&\int d^4x\, \sqrt{-g}\;\Big(\ol j ^{\TT\,\m}\;
\frac{1}{\sl D+m}\;j _\m^\TT
+\ol j .\g\, \phi(j )\Big)\nn\\
&=&
\int d^4x\, \sqrt{-g}\;
\Big(\ol j ^{\m}\;
\frac{1}{\sl D+m}\;j _\m
\nn\\&&\qquad\quad
+\frac{1}{2}\;\ol j .\g\,
\frac{1}{3D^2+\L}\,
\Big[
\frac{2D^2\,(\sl D+m)+\,\L\sl D}{(D^2-m^2)}
+\frac{\L(\sl D-2m)}{(\L+3m^2)}
\Big]\,\g.j 
\Big)\, .
\nn\\
\label{cheese}
\eea
Firstly note that the apparent pole at $D^2=-\L/3$ is spurious; a simple
calculation reveals a vanishing residue there. At the physical pole
$D^2=m^2$ we find in the $\g.j $ sector
\be
{\cal A}(\g.j )\stackrel{D^2\rightarrow m^2}{\sim}
\int d^4x\, \sqrt{-g}\;
\ol j .\g\;
\frac{(m^2+\frac{1}{2}\,\L)\,\sl D+m^3}{3m^2+\L}\;
\frac{1}{D^2-m^2}\;\g.j \, .
\label{formaggio}
\ee   
In the limit where we first go to flat space and thereafter
take the mass to zero, we get
\be
{\cal A}(\g.j )\sim
\frac{1}{3}\;
\int d^4x\;
\ol j .\g\;
\frac{1}{\sl \d}
\;\g.j \, ,
\qquad
(\L\rightarrow 0\mbox{ \underline{then} } 
m\rightarrow 0)\, .
\ee
For continuity, the prefactor should be~1/2 and this is the spin~3/2
version~\cite{Deser:1977ur} of the famous spin~2 discontinuity.
Now take the limit $m\rightarrow0$ first
\be
{\cal A}(\g.j )\sim
\frac{1}{2}\;
\int d^4x\;
\ol j .\g\;
\frac{1}{\sl \d}
\;\g.j \, ,
\qquad
(m\rightarrow 0\mbox{ \underline{then} } 
\L\rightarrow 0)\, .
\ee
The prefactor is now~1/2 and therefore connects smoothly
to the massless flat space result\footnote{See~\cite{Deser:1977ur}, 
a pedagogical
derivation may be found in~\cite{VanNieuwenhuizen:1981ae}.}
\be
{\cal A}=
\int d^4x\;
\ol j ^\m\;
\Big(
\frac{\eta_{\m\n}}{\sl \d}+\g_\m\,\frac{1}{2\,\sl \d}\,\g_\n
\Big)
\;j ^\n\, ,
\qquad
(m\equiv 0)\, .
\label{light}
\ee
As claimed in the introduction, the ``de Sitter'' limit is continuous.

We now turn to the AdS case and ask the question:
When is a spin~3/2 particle massless in an AdS space? The
answer~\cite{Deser:1977uq} 
is \underline{not} at the value of the parameter $m=0$. Rather,
masslessness occurs whenever a gauge invariance appears. The relevant
gauge invariance for spin~3/2 in AdS is under transformations
\be
\delta \psi_\m=\D_\m \e=(D_\m+\frac{1}{2}\,m\g_\m)\,\e \, ,\; 
\mbox{ \underline{with} }\;   m^2=-\L/3 \, ,
\ee 
{\it i.e.}, at the point where the model becomes (linearized) cosmological 
supergravity~\cite{Deser:1977uq}. 
This is obvious from the constraint~\eqn{Bianchi}
which is now  the Bianchi identity of the locally supersymmetric
gauge theory
\be
\D_\m\,\g^{\m\n\r}\D_\n\psi_\r=0 \, ,\;
\mbox{ \underline{with} }\;   m^2=-\L/3 .
\label{iks}
\ee
Further the ``de Sitter'' calculation presented above is singular in this
limit due to the final term in~\eqn{cheese}.
The reason is simply that we took 
$D.j=0$, which is not consistent with the massless Bianchi identity~\eqn{iks}.

The remedy is to impose 
the conservation law of the underlying massless theory
for the current $J_\m$ 
\be
D.J+\frac{1}{2}\,m\,\Big(\frac{\L}{-3m^2}\Big)^\beta\,\g.J=0
\label{vary}
\ee
We have included the parameter $\beta$ since any of the above choices
reduces to the conservation law $\D.j=0$ of the massless theory at $3m^2=-\L$.
However, let us now argue that the choice $\beta=1/2$ is the correct
one physically. In that case
\be
D.J+\frac{1}{2}\,\sqrt{-\L/3}\,\g.J=0
\label{conservation}
\ee
and the external current obeys the conservation law of the massless theory
for {\it any} massive deformation $m$; we hold the background 
cosmological spacetime constant and study the limit 
$m\rightarrow \sqrt{-\L/3}$. Indeed, an external current cannot depend
on the particular dynamical theory to which it couples, but can, of course,
involve the external background geometry.
In other words, equation~\eqn{conservation} is the analogue of
covariant conservation in AdS space for spin~3/2 fields.

A simple example is convincing: Consider the flat space massive
spin~3/2 propagator
\be
S_{\m\n}^F=\frac{1}{\d^2-m^2}\;\Big[
\Big(\eta_{\m\n}-\frac{\d_\m\d_\n}{m^2}\Big)\,(\sl \d -m)
+\frac{1}{3}\,\Big(\frac{\d_\m}{m}-\g_\m\Big)\,
(\sl \d+m)\,
\Big(\frac{\d_\n}{m}-\g_\n\Big)\,
\Big] ,
\ee
and take its expectation between currents satisfying
$\d.\J+\frac{1}{2}\,m\g.\J=0$. The result is
\be
\ol \J^\m S_{\m\n}^F\,\J^\n=
\ol \J^\m\,\frac{1}{\sl \d+m}\,\J_\m\,
+\frac{1}{2}\,\ol \J.\g\,\frac{\sl \d+2m}{\d^2-m^2}\,\g.\J\, .
\label{late}
\ee
This amplitude connects smoothly with the massless result as
$m\rightarrow 0$ since the
coefficient of the gamma-trace terms is~1/2. This does not mean
that there is no Minkowski discontinuity however, since here the current
$\J_\m$ (which transforms under variations of $m$),
is not the conserved one of the underlying massless theory.
(We note in passing, that although the calculation of the exchange 
amplitude for the choice $\beta=0$ in~\eqn{vary} is rather simple, the
result yields no discontinuity since it is the cosmological analogue
of~\eqn{late}).

Our task now is to compute the exchange amplitude for currents
obeying~\eqn{conservation} and mass-parameter {\it detuned} 
from the supersymmetric value
$3m^2=-\L$. Defining
\be
\ol m\equiv \sqrt{-\L/3}\, ,
\ee
we therefore eliminate $D.J$ throughout via the relation~\eqn{conservation},
\be
D.J=-\frac{1}{2}\,\ol m\, \g.J\, .
\label{eliminator}
\ee 
The method follows the above ``de Sitter'' calculation.
Firstly the gamma-trace and $\D_\m$-divergence of the
Rarita--Schwinger
equation
\be
\g^{\m\n\r}\D_\n\psi_\r=J^\m\, ,
\ee
yield
\be
\g.\psi=-\frac{\g.J}{3(m+\ol m)}\, ,\qquad
D.\psi=-\frac{2\sl D+3\ol m}{m+\ol m}\;\g.J\, .
\ee
(We stress that $\ol m\neq m$ has been used to derive the above.)
In turn, the trace and transverse components of the field $\psi_\m(J)$ are
\be
\phi(J)=
\frac{\ol m\,(\sl D+2\ol m)}{6\,(m+\ol m)\,(D^2-\ol m^2)}\,\g.J\, ,\qquad
\chi(J)=-\frac{\sl D+2\ol m}{3\,(m+\ol m)\,(D^2-\ol m^2)}\,\g.J\, .
\label{oil}
\ee
Again, the half-integer Lichnerowicz identities~\eqn{lick} 
imply that a one-particle exchange amplitude takes the form
\be
{\cal A}=
\int d^4x\, \sqrt{-g}\;
\Big(
\ol J^{\TT\m}\;
\frac{1}{\sl D+m}\;
J^\TT_\m
+
\ol J.\g\,\Big[\phi(J)-\frac{1}{2}\,\ol m\,\chi(J)\Big]\,
\Big)
\, .
\label{Marshall}
\ee
where we have applied~\eqn{eliminator} to eliminate $\ol J.\overleftarrow{D}$.
Finally, the transverse--gamma-traceless decomposition  simplifies
by virtue of~\eqn{eliminator}
\be
J_\m^\TT=J_\m-
\Big[
(\g_\m D^2-D_\m\sl D)+\frac{1}{2}\,\ol m\,(\g_\m\sl D-4D_\m)
\Big]\,
\frac{1}{3D^2+\L}\,\g.J\, .
\label{sump}
\ee
Inserting~\eqn{sump} and~\eqn{oil} into the amplitude~\eqn{Marshall}
gives the central result\footnote{To check the details, the identity
$$
\frac{1}{\sl D+m}\,(\g_\m\sl  D-4D_\m )\,\chi=
(\g_\m\sl  D-4D_\m )\,\frac{\sl D-m}{D^2-m^2}\,\chi
-2\,(\g_\m D^2-D_\m\sl D)\,\frac{1}{D^2-m^2}\,\chi
$$
valid for any spinor $\chi$, will be appreciated.}
\bea
{\cal A}
&=&
\int d^4x\, \sqrt{-g}\;
\left(\;\ol J ^{\m}\,
\frac{1}{\sl D+m}\,J _\m\right.
\nn\\
&+&
\frac{1}{3}\;
\ol J .\g\;
\frac{1}{D^2-\ol m^2}\,
\left[
\frac{D^2\,(\sl D+m+\ol m)-\ol m\,([2\ol m-m]\sl D+3\ol m^2-m\ol m)}
{(D^2-m^2)}\right.
\nn\\&&\qquad\qquad\qquad\qquad\qquad\qquad\qquad\qquad\left.\left.
\,+\,\;
\frac{\ol m\,(\sl D +2\ol m)}{m+\ol m}\,
\right]\,\g.J 
\right)\, .
\eea
Again an apparent pole at $D^2=\ol m^2$ is spurious: it is easy to
verify that its residue cancels (a strong check of our algebra).
Observe also that, unlike for~\eqn{cheese}, there is no singularity at
$m=\ol m=\sqrt{-\L/3}$.
At the physical pole
$D^2=m^2$ we find in the $\g.J$ sector
\be
{\cal A}(\g.J )\stackrel{D^2\rightarrow m^2}{\sim}
\int d^4x\, \sqrt{-g}\;
\ol J .\g\;
\frac{(m+2\ol m)\, \sl D+m^2+2m\ol m+3\ol m^2}{3\,(m+\ol m)\,(D^2-m^2)}\;
\g.J
\ee   
Firstly, consider the limit in which we expect a discontinuity, namely
setting the cosmological constant $\L$ to zero
first, and only thereafter the mass $m\rightarrow0$,
\be
{\cal A}(\g.J)\sim
\frac{1}{3}\;
\int d^4x\;
\ol J .\g\;
\frac{1}{\sl \d}
\;\g.J \, ,
\qquad
(\L\rightarrow 0\mbox{ \underline{then} } 
m\rightarrow 0)\, .
\ee
The prefactor should be~1/2 and again we find the spin~3/2
version of the spin~2 discontinuity.

Finally, we show that cosmological supergravity 
is free of any discontinuity. That means taking the (supersymmetric) limit
\be
m\rightarrow \sqrt{-\L/3}\,\Leftrightarrow\, m\rightarrow \ol m\, ,
\ee
in which the theory truly becomes (AdS) massless.
The result is
\be
{\cal A}(\g.J )\sim
\frac{1}{2}\;
\int d^4x\, \sqrt{-g}\;
\ol J .\g\;
\frac{\sl D+2\ol m}{D^2-\ol m^2}
\;\g.J \, ,
\ee
This is the result for the exchange amplitude of cosmological
supergravity. The final step is to take
the flat space limit $\L\rightarrow 0$ which, of course, is smooth
\be
{\cal A}(\g.J )\sim
\frac{1}{2}\;
\int d^4x\;
\ol J .\g\;
\frac{1}{\sl \d}
\;\g.J \, ,
\qquad
(m\rightarrow \sqrt{-\L/3}\mbox{ \underline{then} } 
\L\rightarrow 0)\, .
\ee
The prefactor~1/2 of course matches perfectly with the strictly
massless result~\eqn{light}.

To summarize, we have seen that the spin~3/2 one-particle exchange amplitude
in AdS  connects smoothly to (massless) cosmological
supergravity. This requires the limit $m\rightarrow \sqrt{-\L/3}$
along with the current conservation law $\D(\L).J=0$ of the
underlying massless supergravity theory.
The flat massless Minkowski limit can be achieved continuously
thereafter by taking $\L \rightarrow 0$.

\section*{Acknowledgments}
This work was
supported by the National Science Foundation under grant PHY99-73935.

\end{document}